# Comprehensive Assessment of COVID-19 Impact on Saskatchewan Power System Operations

Nima Safari [1*], George Price[1], C.Y. Chung [2]

[1] Grid Operations Support, SaskPower, Regina, Saskatchewan, Canada

[2] Department of Electrical and Computer Engineering, University of Saskatchewan, Saskatoon, SK S7N 5A9, Canada

*E-mail: nsafari@saskpower.com

**Abstract:** This paper presents lessons learned to date during the Coronavirus Disease 2019 (COVID-19) pandemic from the viewpoint of Saskatchewan power system operations. A load estimation approach is developed to identify how the closures affecting businesses, schools, and other non-critical businesses due to COVID-19 changed the electricity consumption. Furthermore, the impacts of COVID-19 containment measures and re-opening phases on load uncertainty are examined. Changes in CO2 emissions resulting from an increased proportion of renewable energy generation and the change in load pattern are discussed. In addition, the influence of COVID-19 on the balancing authority's power control performance is investigated. Analyses conducted in the paper are based upon data from SaskPower corporation, which is the principal electric utility in Saskatchewan, Canada. Some recommendations for future power system operation and planning are developed.

## 1. Introduction

Coronavirus disease 2019 (COVID-19) is a contagious disease caused by severe acute respiratory syndrome coronavirus 2 [1]. COVID-19 was first recognized in December 2019, in Wuhan, China and, soon after, COVID-19 cases were identified all over the world and the number of infected individuals progressively increased. The first confirmed case in Canada was determined on January 27, 2020. In Canada, travelers or individuals linked to the travelers primarily constituted the initial confirmed COVID-19 cases, which in turn galvanized the Federal Government to invoke the Quarantine Act in mid-March, 2020. As per this Act, travelers, excluding essential workers, are legally required to self-isolate for 14 days upon entering Canada from another country. On March 11, 2020, the World Health Organization (WHO) declared COVID-19 a global pandemic. Correspondingly, various provincial governments announced a State of Emergency (SOE).

By March 22, an SOE had been declared in all Canadian provinces and territories. As a consequence, all schools, universities, and non-essential business centers were closed. International travel was severely restricted and large assemblies were prohibited. Employees of many organizations and companies were asked to work remotely from home. As a side-effect of these socio-economic changes, various industries experienced an unprecedented recession. From February 21 to March 27, the Canadian Crude Index (CCI) tumbled to $8.66 (i.e., the lowest price in five years) [2]. The Alberta Natural Gas price also declined by 10% from February to March 2020 [3]. The stock market price of various multinational industries, which make up a considerable portion of the non-conforming electric loads, plummeted drastically, which can be considered as an indicator of lower manufacturing activities. For instance, on March 13, the stock price for EVRAZ plc had dropped by 41% compared to February 21, and on March 20, Nutrien Ltd. and Mosaic Co. experienced their lowest stock market price in the last five years [4], these companies are significant electricity users in Saskatchewan, Canada. Provincial Government measures, devised to tackle the COVID-19 pandemic, as well as their consequent impact on human activities and industries, resulted in drastic changes in electricity consumption in Saskatchewan. In turn, power system operations have changed, and operators have faced unconventional challenges.

Changes in electricity consumption patterns have been reported in various countries and regions in response to the pandemic. In Italy, electricity consumption in 2020 declined by 37% compared to 2019, and this demand reduction affected the electricity market and power system reliability. The energy price also decreased by approximately 30%, forcing the large thermal generation units (which cannot compete against renewables) to shut down due to higher production costs compared to the market price [5]. In the Brazilian power grid, the lockdowns and isolations resulted in an apparent decline in electricity consumption and changes to the weekly load profile [6]. The impact of various European countries measures to curb the spread of COVID-19 on electricity consumption is examined in [7]. Electricity consumption has noticeably differed during COVID-19, compared to the same time period in 2019. Electricity consumption declined in European countries undergoing lockdowns but grew in other countries with less restrictive measures. Changes in power systems operations in three small power systems—Israel, Estonia, and Finland—is examined in [8]. Compared to

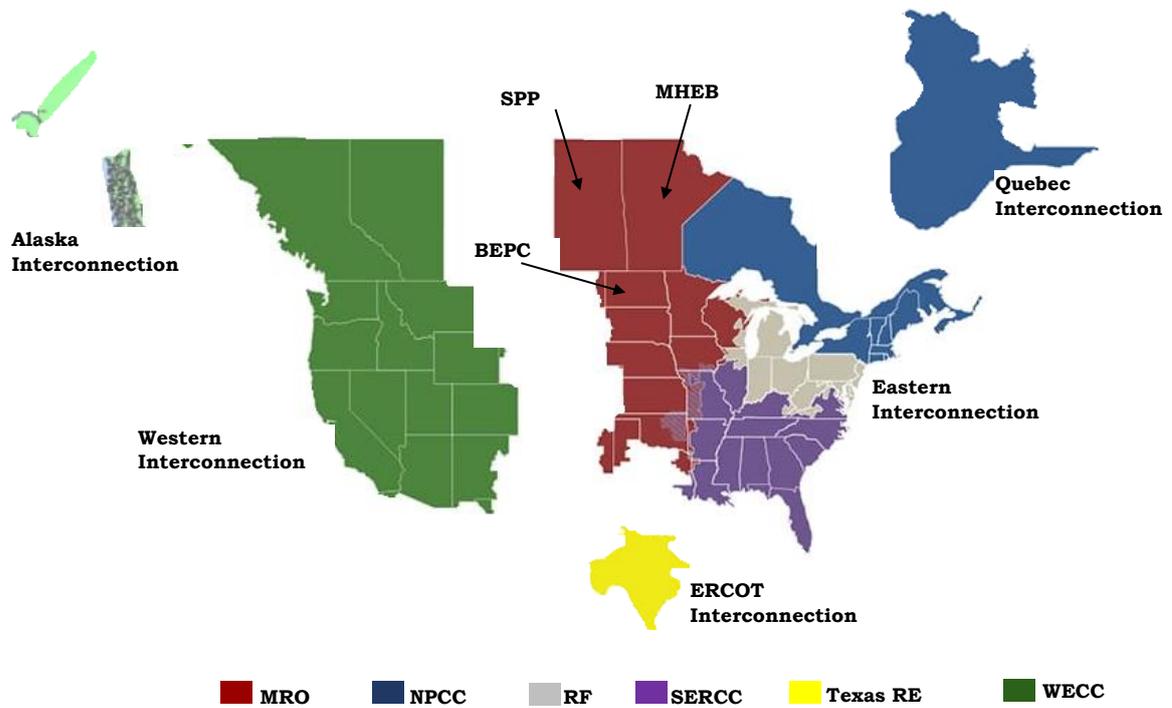

*Fig. 1. North American interconnections, NERC's regional reliability entities, and SPC's neighbors.*

February 2020, the commercial load in both Israel and Estonia dropped more than 15% by April 2020. The weekly averaged load profile also became smoother as time progressed. Overvoltage and frequency stability issues resulting from low loads and high renewable penetration are considered the key problems in these grids.

The experiences shared by various electric utilities show that depending on the power system specifications and governmental measures, utilities may face dissimilar and unprecedented challenges in power system operations; therefore, investigating the impact of the COVID-19 can provide further insights into the effect of a pandemic on electric power systems operation. Moreover, most of the studies were conducted early in the COVID-19 outbreak, and the prolonged influence of the pandemic and response measures have not been examined. To this end, this paper considers approximately six months of historical data (since the start of the COVID-19 pandemic) and scrutinizes Saskatchewan power system operations during the initial months of the SOE and the mandatory business closures and then the subsequent re-opening phases. Saskatchewan power system operations are analyzed based on historical data recorded before and during the health crisis (i.e., January 2015 to September 2020). A weather-based load estimation approach is developed to estimate the changes to the daily load pattern, as well as energy consumption. Different aspects of the load profile are examined, which can be used to define scenarios that need to be considered in planning studies. Additionally, the impact of a modified daily electric load profile, as a result of the SOE, on the accuracy of the load forecasting model is considered. The consequences of prolonged energy consumption disruption on the generation mix, $CO_2$ emissions, and power control performance are discussed. Taking into account the outcomes of this study in planning and operation can facilitate power system resiliency and future preparedness against pandemics.

## 2. Saskatchewan's Power System

To acknowledge the impact of COVID-19 containment measures on Saskatchewan power system operation, some information is provided about North American power systems as well as SaskPower Corporation (SPC)—the primary electric utility in Saskatchewan, Canada.

As shown in Fig. 1, the electrical power grid of North America is mainly a network of five interconnected grids: Eastern Interconnection, Western Interconnection, Texas Interconnection (ERCOT), Quebec Interconnection, and Alaska Interconnection. The North American Electric Reliability Corporation (NERC)—the electric reliability organization for North America—oversees six regional reliability entities to ensure the reliability of the North America Electric Power System [9]: Northeast Power Coordinating Council (NPCC), Midwest Reliability Organization (MRO), SERC Reliability Corporation (SERC), Reliability First (RF), Texas Reliability Entity (Texas RE), and Western Electricity Coordinating Council (WECC). Reginal reliability entities are responsible for ensuring the compliance of the entities that own, operate, or use the interconnected grid.

SPC is part of the Eastern Interconnection and a subregion within the MRO. SPC serves as a balancing authority, transmission operator, and reliability coordinator that plans and operates generation and transmission systems such that the demand and resource balance is maintained within the balancing authority area and supports the interconnection frequency in real-time. As can be observed from Fig. 1, SPC is the most western region of the Eastern Interconnection and is connected to the eastern region of the interconnection through the Manitoba Hydro Electric Board (MHEB) with five alternative currents (AC) tie-lines. It is also connected to the southern part of the interconnection (i.e., the US) via the Basin Electric Power Cooperative (BEPC) grid by means of a phase shifter transformer. SPC connects the Eastern Interconnection to the Western Interconnection and WECC with a direct current (DC) interconnection to Alberta, Canada.

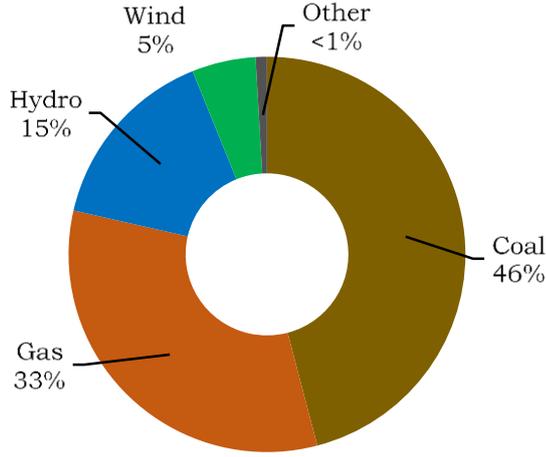

*Fig. 2. SPC installed generation capacity mix.*

SPC's total installed generation capacity is about 4850 MW. Most of the generation facilities are owned by SPC, with the exception of a few independent power producers that are mainly gas-fired generation facilities. The installed generation mix in Saskatchewan is illustrated in Fig. 2. SPC's maximum observed electricity demand in summer (i.e., May to October) and winter (November to April) are 3523 and 3792 MW, respectively. Compared to the Eastern Interconnection, SPC's generation and load are relatively small. Therefore, due to SPC's AC tie-lines with the rest of the Eastern Interconnection, SPC benefits from a high system inertia.

## 3. Provincial COVID-19 Containment Measures

Identifying the COVID-19 control measures that have disturbed the energy consumption pattern is essential to understand the impact of the pandemic on Saskatchewan power system operations. A summary of the Provincial Government responses is provided below.

On March 18, 2020, the Government of Saskatchewan declared an SOE, which was followed by the closure of all businesses except critical public services and essential business services. Critical public services and essential business services are defined by the Government as services and businesses the closure of which would imperil residents' health and safety, or that are providing residents with essential needs. All gatherings were limited to 10 people, and all activities that require larger assemblies were cancelled or became virtual. As per Governmental guidelines, many companies had their employees to work from their own homes.

On April 23, 2020, the Government of Saskatchewan released its re-opening plan, which consisted of five phases. In the first phase, starting May 4, some minor businesses, including previously restricted medical services, were re-opened, but the size of gatherings was still limited to 10 people. On May 19, retail and personal care services were re-opened in the second phase, but the gathering size limit remained at 10. In the third phase of re-opening that began June 8, restaurants, fitness centers, places of worship, and childcare facilities recommenced their activities, and the allowable size of indoor and outdoor assemblages increased to 15 and 30, respectively. In the fourth phase of re-opening, starting June 22, most restrictions on continuity of services and businesses were lifted. The fifth phase (date forthcoming) is a long-term plan to remove the restriction on the size of the public gatherings. Along with the Government re-opening plan, many companies have established their re-integration plans with several phases to return staffs who are working remotely to the workplace.

As elaborated in the next section, the SOE declaration and COVID-19 control measures initially resulted in a reduction in energy consumption, as well as noticeable changes in the energy consumption pattern. However, in response to the re-opening phases and re-integrations, energy consumption has gradually returned to levels closer to the normal.

## 4. Impact of COVID-19 on Electricity Consumption

To examine the consequences of the COVID-19 on electricity consumption, in this section, first, a demand estimation approach is developed that approximates the electricity demand based on historical load patterns and climatological information during SOE, in the absence of the COVID-19 impacts. Comparing the estimated energy consumption with the actual one highlights the impacts of the COVID-19 on the electricity demand.

### 4.1. An Electricity Demand Estimation Approach

Hereafter the hourly averaged system load time series is denoted by $\{y_t\}_{t=T_s}^{N \cdot T_s}$, where $T_s$ is the sample time. In this paper, $T_s = 1\ h$. The electricity demand during the COVID-19 pandemic is estimated using the following linear model.

$$\hat{y}_t = trend_{year_t} + \boldsymbol{a}_t \times \boldsymbol{b}, \qquad (1)$$

$$\begin{aligned}
\boldsymbol{a}_t = [&WeekDay_t, Month_t, \\
& Hour_t, WeekdayHour_t \\
& Weekend_t, Holiday_t, \\
& Temp_t^{Max} \cdot Holiday_t, Temp_t^{Min} \cdot Holiday_t, \\
& WindChill_t \cdot Hour_t \cdot WindChillSeason_t, \\
& HeatIndex_t \cdot Hour_t \cdot HeatSeason_t, \\
& Temp_t \cdot Hour_t, Temp_t^2 \cdot Hour_t, 1],
\end{aligned} \qquad (2)$$

where $trend_{year_t}$ pertains to the annual trend of the load for time $t$. Throughout a year, $trend_{year_t}$ has a constant quantity. $\boldsymbol{WeekDay_t}$, $\boldsymbol{Month_t}$, $\boldsymbol{Hour_t}$, and $WeekdayHour_t$ are one-hot vectors with $1 \times 7$, $1 \times 12$, $1 \times 24$, and $1 \times 168$ vector, respectively. $Weekend_t$ is a binary variable, which is 1 when time $t$ refers to a time on a weekend and 0 otherwise. In (2), $Holiday_t$ is a $1 \times 12$ one-hot vector, where the $i$th element of the vector becomes 1 if time $t$ is related to a federal or provincial statutory holiday in the $i$ th month. $WindChillSeason_t$ and $HeatSeason_t$ are binary variables that categorize the months of the year into the cold season (i.e., months with wind chill) and hot season (i.e., months with high heat index). In (2), $Temp_t^{Max}$ and $Temp_t^{Min}$ are the maximum and minimum daily temperatures, respectively. $Temp_t$, $WindChill_t$, and $HeatIndex_t$ pertain to ambient temperature, wind chill, and heat index, respectively. $\boldsymbol{b}$, in (1), is an unknown column vector. $trend_{year_t}$, in (1), is obtained by least square approach without consideration of the second term of Eq. 1 (i.e., $\boldsymbol{a}_t \times \boldsymbol{b}$). After acquiring $trend_{year_t}$, $\boldsymbol{b}$, in (1), is acquired using the least-squares approach. Climatological data for this study are obtained from [10], and the wind chill and heat index are calculated based on [11] and [12], respectively. It is worth noting that other feature candidates including various lags of climatological data are tested as the input features, however, no improvement in the load estimation was observed.

### 4.2. Evaluation Metrics

Mean Squared Error (MSE) is employed to examine the accuracy of the load estimation and assess the load consumption changes resulting from the COVID-19 pandemic. The MSE is calculated as follows [13]:

$$MSE = Var + Bias^2, \quad (3)$$

where $Var$ and $Bias$ are the variance and average of error time series (i.e., $\{y_t - \hat{y}_t\}_{t=T_s}^{N \cdot T_s}$), respectively.

The daily Energy Variation Index ($EVI$) is another metric to assess further the trend of energy consumption and measures the percentage of the changes in energy consumption with respect to the estimated daily energy. The $EVI$ for the $d$th day is defined as follows:

$$EVI_d(\%) = \frac{\sum_{t=t'}^{t''} \hat{y}_t - y_t}{\sum_{t=t'}^{t''} \hat{y}_t} \times 100, \quad (4)$$

$$t' = 24 \cdot (d-1) + 1, \quad (5)$$

$$t'' = 24 \cdot d. \quad (6)$$

### 4.3. Load Estimation

An easy way to comply with the requirements stated in the Author Guide [1] is to use this document as a template and simply type your text into it. PDF files are also accepted, so long as they follow the same style.

To investigate the performance of the load estimation approach prior to COVID-19 and assess the impact of the COVID-19 containment measures on electricity consumption, three scenarios are considered. Table 1 summarizes the time window of each scenario. In *Scenario I*, the hourly averaged load from March 1, 2017 to March 18, 2020 (i.e., prior to the declaration of a provincial SOE) is estimated. The performance evaluation of the load estimation for *Scenario I* is used to examine the accuracy and unbiasedness of the load estimation. In the load estimation process, the model is updated every 167 days (i.e., the number of days from the SOE declaration to September 1, 2020) using the recent 26 months dataset. Updating the model every 167 days ensures the historical load data do not influence the load estimation. *Scenarios II* and *III* represent the estimated hourly averaged load from March 18 to September 1 in 2019 and 2020, respectively. These scenarios allow for a comparison of the load estimation performance during similar days in different years, where the latter reflects dates after the SOE declaration.

The results of $MSE$ and its components (i.e., $Var$ and $Bias$) from the estimation of various scenarios are presented in Table 1. $MSE$ values for *Scenario III*, compared to the ones for *Scenarios I* and *II* is substantially higher which is due to its evidently higher $Bias$ value as the result of lower energy consumption compared to expected energy consumption during the time interval associated to *Scenario III*. Negative $Bias$ corresponds to an overall overestimation (i.e., estimated loads are higher than actual loads), while positive values refer to a general underestimation (i.e., i.e., estimated loads are lower than actual loads). This table shows the load estimation approach results in a relatively unbiased prediction for *Scenarios I* and *II*, but a tangible overestimation in *Scenario III*. To compare the $Bias$ values for *Scenarios I* and *II* with the one for *Scenario III*, it is assumed that the load estimation error for each hour of all scenarios can be modeled as independent random variables each with equal variance. It can be shown that there is abnormal load behavior in *Scenario III*, compared to reference Scenario (i.e., *Scenarios I* and *II*), if the following condition holds:

**Table 1** Comparisons between the actual load and estimated load for various scenarios.

| Scenarios | Definitions | $MSE$ | $Var$ | $Bias^2$ | $Bias$ |
|---|---|---|---|---|---|
| *Scenario I* | March 1, 2017 to March 18, 2020 | $1.28 \times 10^4$ | $1.26 \times 10^4$ | $0.02 \times 10^4$ | $-13.04$ |
| *Scenario II* | March 18, 2019 to September 1, 2019 | $0.88 \times 10^4$ | $0.85 \times 10^4$ | $0.03 \times 10^4$ | $16.60$ |
| *Scenario III* | March 18, 2020 to September 1, 2020 | $3.15 \times 10^4$ | $1.53 \times 10^4$ | $1.62 \times 10^4$ | $-127.28$ |

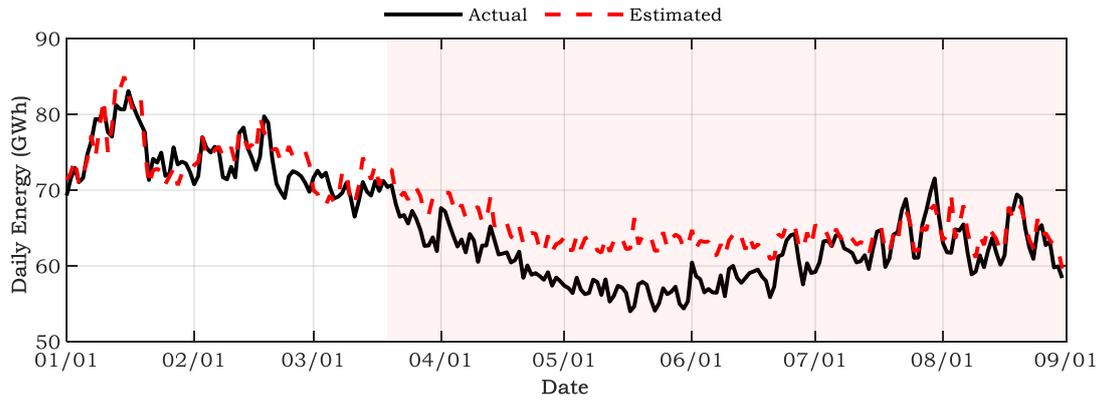

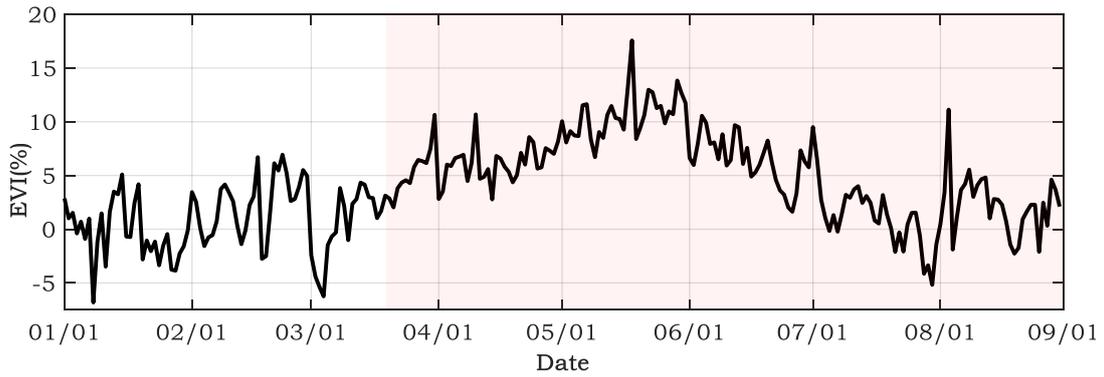

***Fig. 3.*** *Comparison of the actual and estimated hourly averaged load for (**A**) May 2019 and (**B**) May 2020.*

$$\left|Bias_{SC_{Ref}} - Bias_{SC_{III}}\right| \gg max\left(\sqrt{\frac{Var_{SC_{Ref}}}{N_{SC_{Ref}}}}, \sqrt{\frac{Var_{SC_{III}}}{N_{SC_{III}}}}\right) \quad (7)$$

where sub-index $SC_{Ref}$ refers to *Scenarios II* or *III*, and the sub-index $SC_{III}$ is associated with *Scenario II*. Based on Table 1, the Eq. 7 inequality is satisfied, i.e., 114.24 ≫ 1.95 if $SC_{Ref}$ refers to *Scenario I* and 143.88 ≫ 1.95 if $SC_{Ref}$ refers to *Scenario II*. Therefore, the observed *Bias* in *Scenario III* is substantially high which is an indicator of anomalies and can be associated with COVID-19 as discussed further in the following.

Fig. 3a shows actual and estimated daily energy consumption from January 1 to September 1, 2020. Prior to the SOE declaration, the estimated daily energy consumption is very close to the energy consumption; however, since late February 2020, the estimated energy values have been higher, which may be rooted in the historical oil price drop and the subsequent reduced industrial activities. Following the declaration of the SOE in Saskatchewan, electric energy demand declined steadily, with the highest reduction in energy consumption observed in May 2020. However, since mid-July 2020, the difference between actual and estimated daily energy consumption has progressively decreased. To further clarify the difference between the estimated and actual daily energy consumption, Fig. 3b presents the *EVI* for January 1 to September 1, 2020. The *EVI* prior to COVID-19 is mostly limited within the -5% to 5% interval—except in late February when the oil price crashed. However, the *EVI* spiked to more than 10% after the SOE declaration until mid-July, when the fourth phase of re-opening in Saskatchewan started, i.e., when shopping malls, recreational centers, gym facilities, restaurants, small businesses, movie theatres, etc., were re-opened with reduced capacity.

As mentioned above, SPC observed the highest energy consumption drop in terms of *EVI* in May 2020. To further investigate the load behavior, the estimated and actual hourly averaged load values for May 2019 and May 2020 are presented in Fig. 4. Fig. 4a shows the estimation approach can determine the load for May 2019 with high accuracy, except in the first few days of the month that feature non-conforming loads due to planned and unplanned outages. On the other hand, Fig. 4b shows the estimated load for May 2020 is markedly higher than the actual load. Further investigation revealed a statistically significant correlation between maximum daily temperature and *EVI* observed in May 2020 (Pearson correlation 0.63, $p<0.0001$). Fig. 5 shows the *EVI* and maximum daily temperature associated with each day of May 2020. The large differences between the estimated and actual loads mostly occurred on days with high temperatures, when cooling loads needed to be utilized. However, due to the closure of many indoor public areas, teleworking plans, and cancellation of public

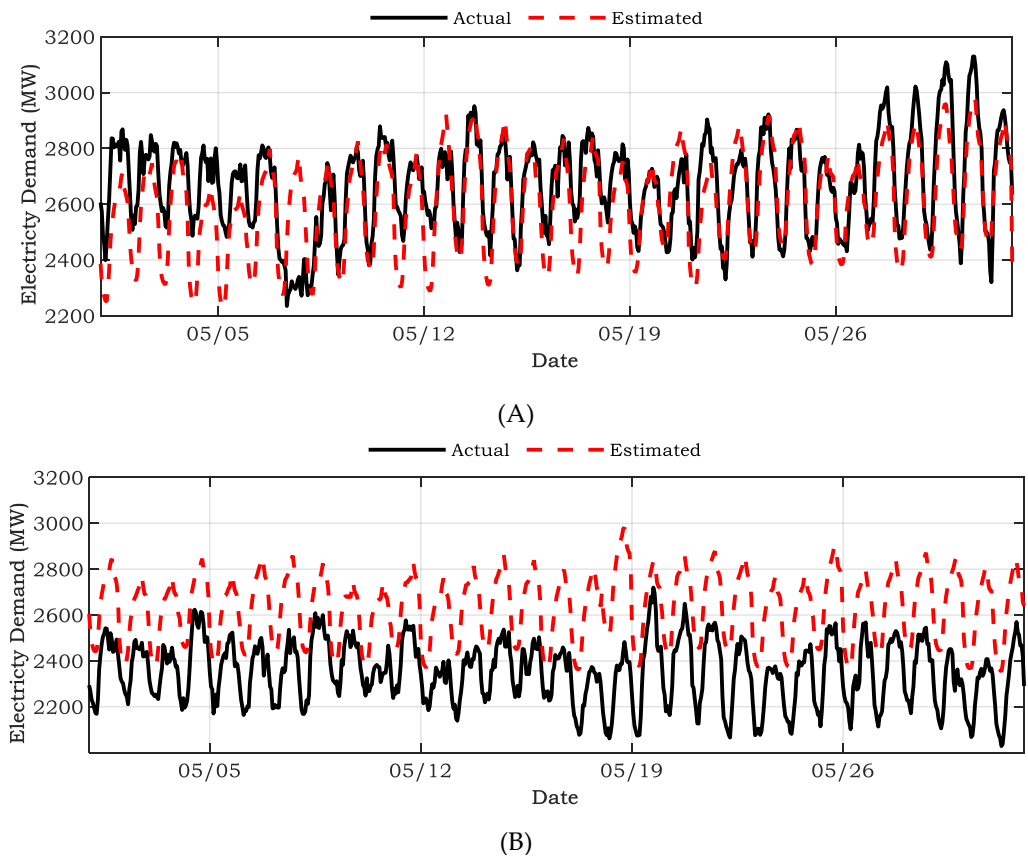

*Fig. 4. Comparison of the actual and estimated hourly averaged load for (A) May 2019 and (B) May 2020.*

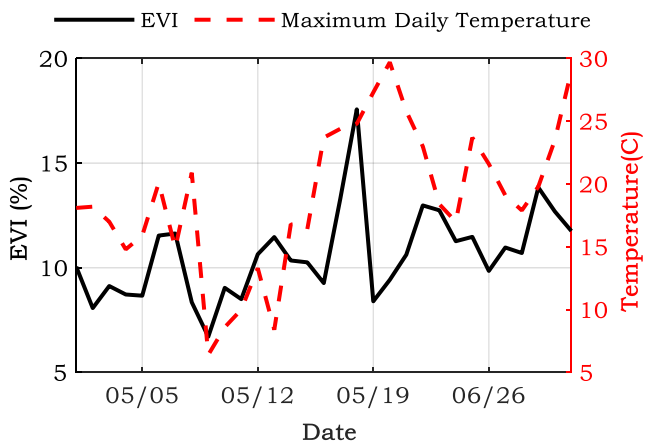

*Fig. 5. EVI and maximum daily temperature in May 2020.*

events, the cooling loads were substantially reduced compared to 2019.

To investigate the accumulated energy consumption reduction since the SOE declaration, the accumulated difference between the estimated and actual load for 2020 is shown in Fig. 6. The accumulated difference for the last three years within a similar period is also depicted and provides insights about the performance of load estimation in past years. The figure shows the accumulated difference between the estimated and actual load for 2017-2019 fluctuates and is relatively small, yet for 2020 grew steadily but has been approximately flat since mid-July. Overall, the 2020 data reflect an approximately 510 GWh reduction in energy consumption, but also that the impact of COVID-19 on energy consumption is diminishing.

From the declaration of the SOE to mid-July 2020, the hourly ramp-up and ramp-down of the daily load profile slightly declined, thereby flattening the daily load profile. Figs. 7 and 8 respectively present box-and-whisker diagrams of the maximum hourly averaged ramp-up and ramp-down that occurred daily from March 18 to July 15 in various years. Approximately 75% of the maximum daily ramp-ups and ramp-downs in 2020 are lower than the median of maximum daily ramp-up and ramp-downs that occurred in the previous five years.

Overall, SPC has observed substantial changes in the electricity load consumption pattern. From March 18 to mid-July 2020, energy consumption dropped. The most noticeable energy consumption decline occurred in May when the ambient temperature was as high as 30 ℃. The hourly average load variability also shrunk; and consequently, the daily load profile became smoother. Since the fourth phase of re-opening began in mid-July, the actual load consumption pattern has begun to resemble the expected pattern more closely. This indicates the energy reduction impact of COVID-19 has declined.

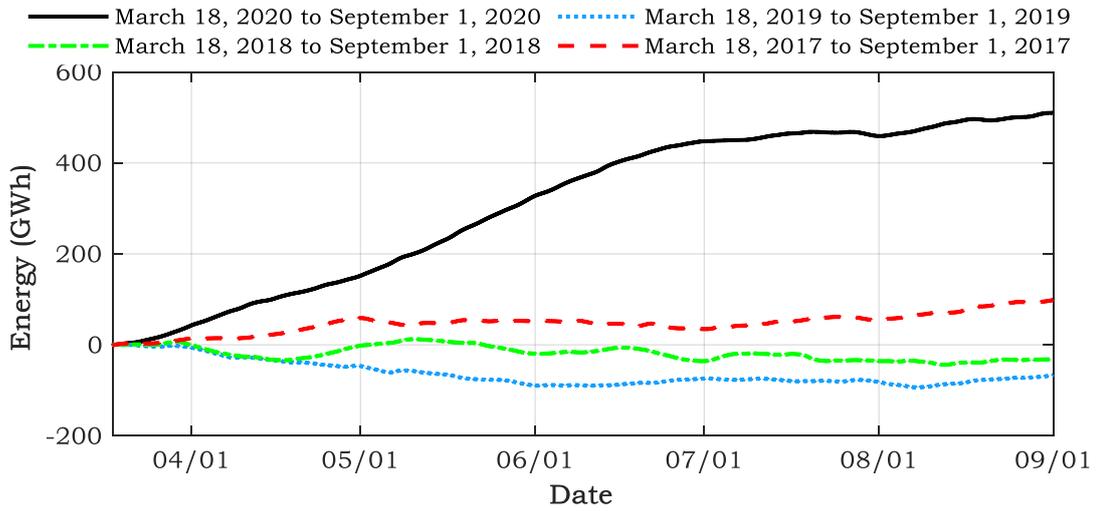

*Fig. 6. Accumulated difference between estimated and actual electricity consumption between March 18 and September 1 for different years.*

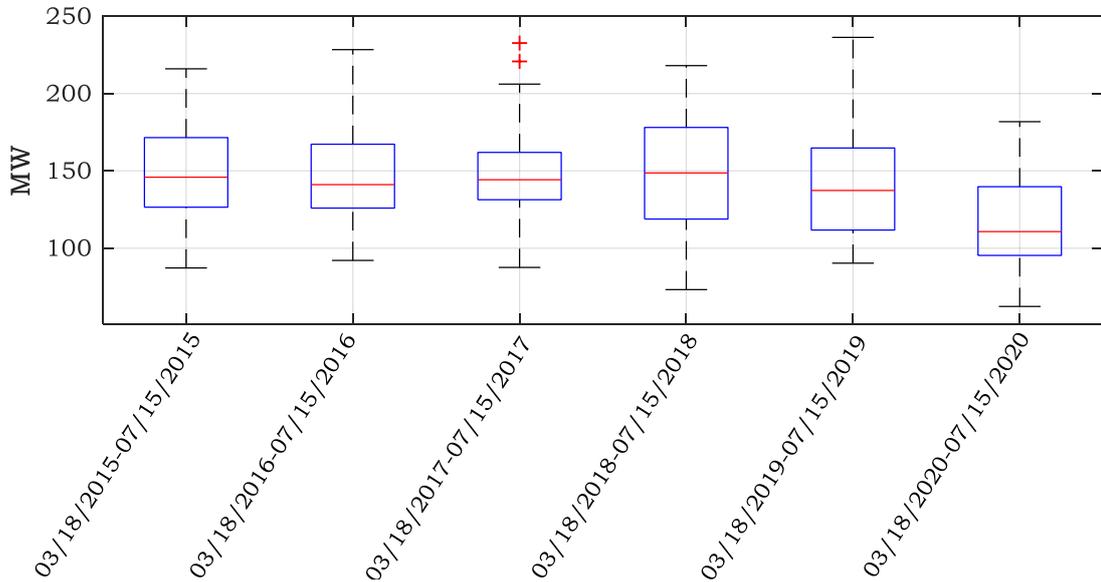

*Fig. 7. Box-and-whisker diagram of the maximum hourly averaged ramp-up between March 18 and July 15 for different years.*

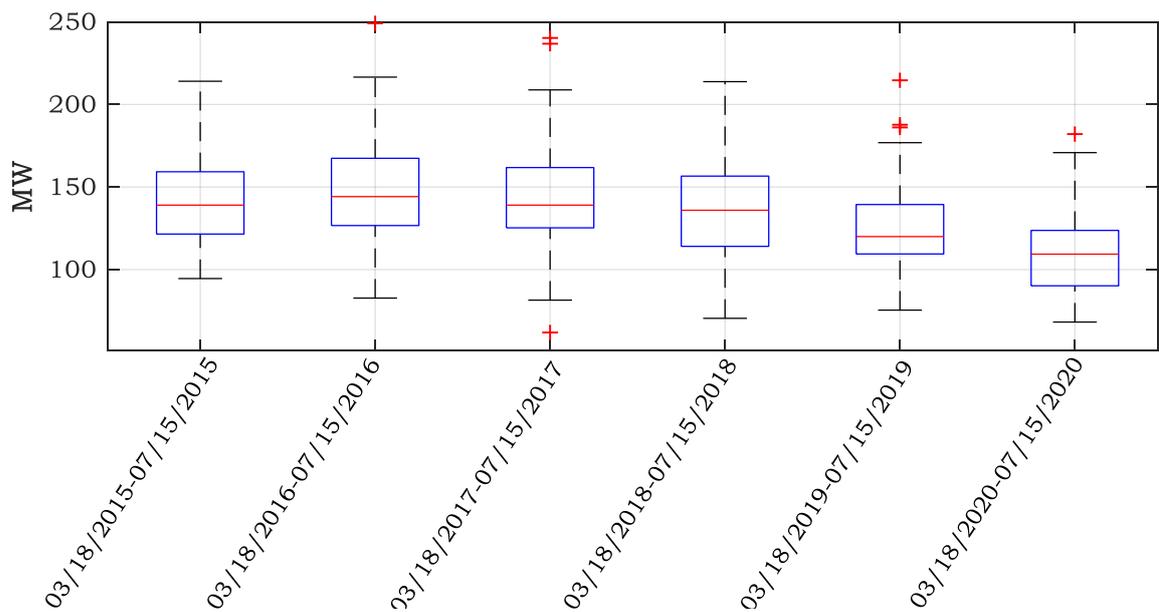

*Fig. 8. Box-and-whisker diagram of the maximum hourly averaged ramp-down between March 18 and July 15 for different years.*

**Table 2** Day-ahead load forecasting performance for from March to September of 2019 and 2020.

| Year | 2019 | | 2020 | | Relative Error (%) | |
|---|---|---|---|---|---|---|
| Month | $MAE$ (MW) | $MAPE$ (%) | $MAE$ (MW) | $MAPE$ (%) | $MAE$ | $MAPE$ |
| March | 47.52 | 1.61 | 80.23 | 2.86 | 68.83 | 77.64 |
| April | 50.05 | 1.86 | 66.29 | 2.60 | 32.45 | 39.78 |
| May | 65.28 | 2.51 | 66.16 | 2.81 | 1.35 | 11.59 |
| June | 50.30 | 1.87 | 59.15 | 2.37 | 17.59 | 26.74 |
| July | 56.87 | 2.08 | 75.76 | 2.80 | 33.22 | 34.62 |
| August | 47.60 | 1.80 | 66.01 | 2.56 | 38.68 | 42.22 |

## 5. Impact of COVID-19 on Load Uncertainty

SPC purchases its load forecasting services from a third-party vendor. The load forecast takes into account weather forecast information as well as the historical load. The day-ahead load forecast is used in the day-ahead unit commitment and economic dispatch. The electricity consumption changes discussed in Section 4 have resulted in increased load forecasting inaccuracy due to a lack of adequate historical load data with a similar pattern. Mean Absolute Error (MAE) and Mean Absolute Percentage Error (MAPE) are used to evaluate the performance of the day-ahead prediction of electricity load for March to September in both 2019 and 2020. MAE and MAPE are calculated as follows:

$$MAE = \frac{1}{N}\sum_{t=1}^{N}|y_t - \hat{y}_t|, \qquad (8)$$

$$MAPE(\%) = \frac{1}{N}\sum_{t=1}^{N}\frac{|y_t - \hat{y}_t|}{y_t} \times 100, \qquad (9)$$

where $y_t$ and $\hat{y}_t$ are the actual and forecasted load values related to the $t$th observation, respectively. $N$ is the total number of observations.

Table 2 summarizes the load forecast performance assessment for various months in 2019 and 2020. The last two columns in this table present the relative $MAE$ and $MAPE$ for 2020 with respect to 2019, where positive quantities refer to deterioration of the forecast in 2020 compared to 2019 for similar months. Compared to the load forecast performance in March 2019, the load forecast in March 2020 significantly deteriorated as the $MAE$ and $MAPE$ increased by 68.83% and 77.64%, respectively. The accuracy of the load forecast gradually recovered in the ensuing months, and in May the load forecast performance was very similar to the previous year. However, since June and with electricity consumption gradually returning to normal, the performance of the load forecast has steadily deteriorated. As noted in Section 4, during the third and fourth re-opening phases, the electricity consumption began to return to normal, but the system load historical data prior to the re-opening phases were considerably lower than the estimated loads. These disturbed historical data might affect the load forecast model.

## 6. Impact of COVID-19 on Fuel, CO2 Emissions, and Generation

The reduced electricity consumption and changes in the daily load profile result in changes to the generation mix. As an example, the pie charts in Fig. 9 show the generation mix in May for 2019 and 2020. This figure shows the penetration of the hydro and wind markedly increased in 2020 versus 2019. Specifically, wind and hydro generation constitute 15% of the total SPC generation in May 2019, but this increased to 32% in May 2020 (i.e., more than 100% increase in their penetration to the power system). Such increased penetration of hydro and wind is associated with the higher runoff flow and substantially lower load in May 2020 compared to May 2019, as discussed in Section 4. Fig. 10 illustrates the highest renewable penetration observed every year since 2015, with the highest percent penetration occurring in 2020. The highest renewable penetration SPC has ever observed was occurred at 8:00 a.m. on May 31, 2020.

The higher share of renewable energy generation contributing to meeting the electricity demand resulted in a reduction in CO2 emissions. Compared to March 18 to September 1, 2019, the same date ranging in 2020 saw CO2 emissions caused by the coal-fired generation fleet drop by more than 25% and by the gas-fired generation fleet drop by 6%. Therefore, thanks to the measures taken to control the COVID-19 pandemic, SPC generation was more environmentally friendly.

With the disruption in the generation mix and reduced load during the COVID-19 pandemic, the real power control performance was also affected. Some of the large fossil fuel-fired generation units could not be dispatched even to their minimum stable generation level and were kept on cold standby mode. This is largely attributed to the lower baseload. On the other hand, the fast response small-scale generation units, which are able to operate in automatic generation control (AGC) mode, were committed and dispatched close to their full available capacity and beyond the AGC mode region. Therefore, they were able to provide very limited regulation services, especially regulation up service. Moreover, Saskatchewan experienced a unique long-lasting spring runoff in which the reservoirs were fully utilized, and a large amount of excess water flowed down the spillways, while most of the hydro generators were generating close to their maximum capacity. Such high runoff flow restricted the hydropower generators' capability

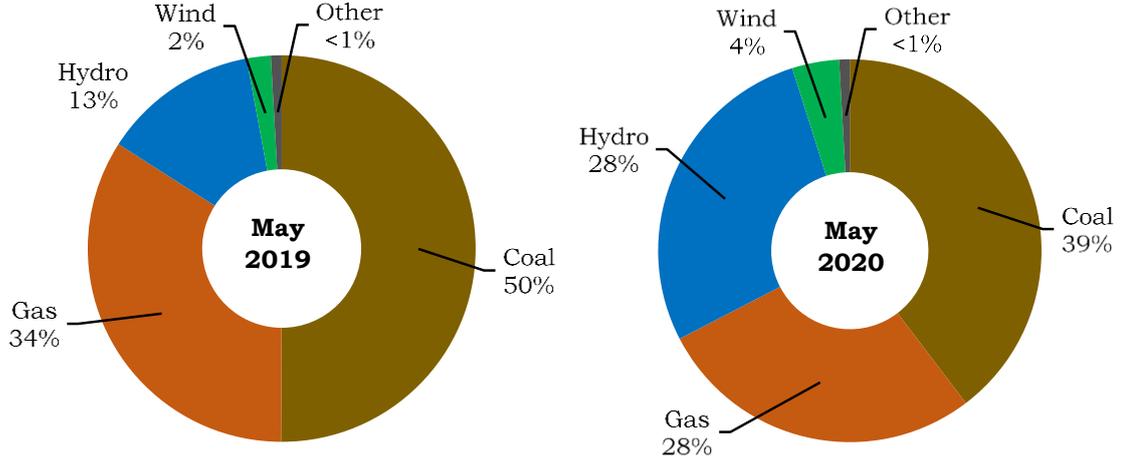

*Fig. 9. Generation mix by fuel in 2019 and 2020 for May.*

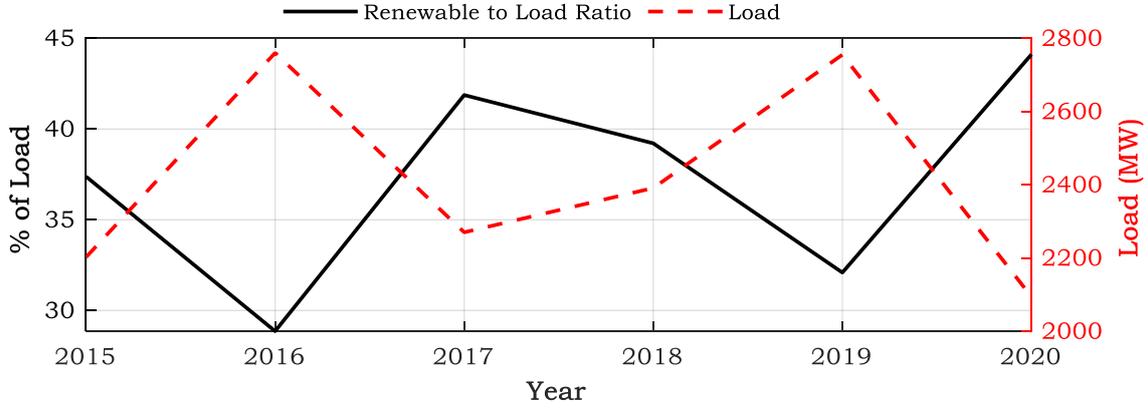

*Fig. 10. Maximum penetration of renewable generation in various years and the load at the hour of maximum penetration.*

to provide down-room regulation reserve.

Due to the aforementioned reasons, the efficiency of power control performance declined; however, SPC's power control performance was still maintained significantly above the NERC requirements. To demonstrate this, Control Performance Standard 1 (CPS1) [14] is calculated for various months. The monthly CPS1 is calculated as follows:

$$CPS1(\%) = (2 - CF) \times 100, \quad (10)$$

$$CF = \frac{CF_{Month}}{(\epsilon 1_I)^2}, \quad (11)$$

$$CF_{Month} = \frac{1}{N_{Month}} \sum_{n=1}^{N_{Month}} CF_{minute}^t, \quad (12)$$

$$CF_{minute}^t = \frac{RACE_{minute}^t}{-10B} \times \Delta F_{minute}^t, \quad (13)$$

where $\epsilon 1_I$ is the interconnection constant, and for the Easter Interconnection is 0.018 Hz. In (12), $N_{Month}$ is the number of minutes in a month. In (13), $B$ is the frequency bias setting, which varies annually for different balancing authorities. $\Delta F_{minute}^t$ is the clock-minute average of frequency deviation from the target system frequency for the $t$th minute and $RACE_{minute}^t$ is the clock-minute average of reporting area control error for the $t$th minute and is calculated as follows [15]:

$$RACE = \big(NI_A(t) - NI_S(t)\big) - 10B\big(F_A(t) - F_S(t)\big), \quad (14)$$

where $NI_A(t)$ and $NI_S(t)$ are the actual and scheduled net interchange at time $t$, respectively. $B$ is the frequency bias setting, expressed as $\frac{MW}{0.1Hz}$. This value for SPC is -41.9 $\frac{MW}{0.1Hz}$ [16]. In (14), $F_A(t)$ pertains to the actual interconnection frequency at time $t$, while $F_S(t)$ is the scheduled frequency.

Table 3 summarizes the relative values of CPS1 for March to September in 2020 compared to 2019. Positive quantities show that, from the CPS1 perspective, the power control performance in 2020 was worse than in 2019, while negative quantities represent a relative improvement in CPS1. Table 3 shows that, in terms of the CPS1 metric, the power control performance has been worse in most months since the SOE declaration, with the worst CPS1 score is related to May; however, from June onward the deterioration in CPS1 score has recovered. Also, note that the CPS1 score for some of the months in 2020 is better than these in 2019; this is attributed in part to the Chinook Combined-Cycle Generation facility—one of the main generation units with a wide range of AGC operation—being out of service.

## 7. Discussion

This study examined various impacts of the socio-economic restrictions resulting from the COVID-19 pandemic. Prolonged COVID-19 containment measures substantially reduced electricity consumption in Saskatchewan. The baseload decreased, which brought about more sustainable power system operations by increasing the penetration of renewable energy, lowering $CO_2$ emissions, and shutting down large-scale fossil fuel-fired generation units. However, the reduced load led to some power system operations challenges. Shutting down some of the fossil fuel-fired generation units meant the available regulating reserve was markedly reduced, which led to a lower CPS1 score in March-June 2020 compared to the same months in 2019. This high renewable penetration provides power systems operators and planners with a unique opportunity to experimentally perceive the challenges of future power systems with increased proliferation of renewables.

Furthermore, as a consequence of COVID-19 socio-economic impacts, system load uncertainty increased as the load forecast model became less accurate. Therefore, in day-ahead and real-time operation planning, more operating reserves should be available to compensate for load forecast errors. No overvoltage or frequency issues were identified, which is attributed to the reduced load as a result of the COVID-19 pandemic.

This study provides valuable insights that can be used in various future studies. For example, a resilient power system must consider scenarios with prolonged electricity consumption reduction in both generation and transmission expansion planning. In this way, the specifications of future generation units (e.g., minimum stable generation level, AGC region, etc.) are identified such that the power system can reliably and economically operate in the case of such scenarios. Correspondingly, from the transmission expansion viewpoint, future expansions must consider adequate flexibility to alleviate issues that can be a consequence of lower electricity demand. From the power system operations perspective, developing load forecasting tools that consider socio-economic factors (i.e., business closures due to pandemics, re-opening and re-integration plans, stock market, etc.) is crucial.

## 8. Conclusion

Socio-economic issues related to the COVID-19 pandemic affected Saskatchewan's electricity consumption. This study demonstrates how provincial measures with respect to the COVID-19 changed the daily load profile and resulted in lower energy consumption. A consequence of the re-opening phases and re-integration processes employed by many organizations and companies diminished the impact of COVID-19 on electricity consumption. Due to the disruption in the electricity demand, the accuracy of the load forecasting tool declined and therefore, more operating reserve was required to handle the load uncertainty. Furthermore, the investigations reveal Saskatchewan experienced its highest ever renewable energy penetration thanks to the lower load during the COVID-19 business closures. The lower load and higher share of renewable energy resulted in drastic changes to the generation mix. Accordingly, $CO_2$ emissions substantially declined in March-September 2020 compared to a similar period in 2019. Furthermore, power control performance declined as a consequence of generation mix changes.

The outcomes presented in this paper provide valuable perceptions of how a prolonged SOE declaration can affect power system operations. The insights will assist planners and operators in ensuring the preparedness of the power system for reliable operation under such circumstances.

## 9. Acknowledgment

The authors gratefully acknowledge Mr. Cordell Wrishko and Ms. Lan Cui for their insightful comments during the preparation of this study.